\documentclass[a4paper]{article}

\usepackage{INTERSPEECH2021}
\usepackage{amsmath,graphicx}
\usepackage{multirow}
\usepackage{amsmath, amssymb}
\usepackage{booktabs}
\usepackage[normalem]{ulem}
\usepackage{color}
\newcommand{\red}[1]{\textcolor{black}{#1}}

\newcommand{\shita}{\mathbin{\rotatebox[origin=c]{-90}{$\curvearrowright$}}}
\newcommand{\ue}{\mathbin{\rotatebox[origin=c]{-90}{$\curvearrowleft$}}}

\title{Acoustic Event Detection with Classifier Chains}
\name{Tatsuya Komatsu$^{1}$, Shinji Watanabe$^{2}$, Koichi Miyazaki$^{3}$, Tomoki Hayashi$^{3,4}$}
\address{$^{1}$LINE Corporation, Japan\\$^{2}$Carnegie Mellon University, USA\\ $^{3}$Nagoya University, Japan\\$^{4}$Human Dataware Lab. Co., Ltd., Japan}
\email{}

\begin{document}
\maketitle
\begin{abstract}
This paper proposes acoustic event detection (AED) with classifier chains, a new classifier based on the probabilistic chain rule.
The proposed AED with classifier chains consists of a gated recurrent unit and performs iterative \red{binary detection} of each event one by one.
In each iteration, the event's activity is estimated and used to condition the next \red{output} based on the probabilistic chain rule to form classifier chains.
Therefore, the proposed method can handle the interdependence among events upon classification, \red{while} the conventional AED methods with multiple binary classifiers with a linear layer and sigmoid function have placed an assumption of conditional independence.
In the experiments with a real-recording dataset, the proposed method demonstrates its superior AED performance to a relative 14.80\% improvement compared to a convolutional recurrent neural network baseline system with the multiple binary classifiers.
\end{abstract}
\noindent\textbf{Index Terms}:
acoustic event detection, multi-label classification, chain rule, classifier chains

\section{Introduction}
Acoustic event detection (AED) is a technology for automatically detecting and recognizing the wide variety of sounds around us and understanding the environment and situation where the sounds have been recorded. 
This technology can be employed for diverse applications, including monitoring and surveillance~\cite{stork2012audio, imoto2013user,radhakrishnan2005audio, clavel2005events}.

AED is a task of labeling semantic events and marking their temporal location and duration in a \red{given audio} signal. Furthermore, it assumes that events can co-occur in time, i.e., AED is a multi-label classification problem.
Hence, some approaches employ source separation techniques, for example, non-negative matrix factorization~\cite{dikmen2013sound, komatsu2016acoustic}.
With increasing data set sizes, neural network-based multi-label classification models have received much attention, e.g., using convolutional neural networks (CNNs)~\cite{gorin2016dcase} and long short-term memory (LSTM)~\cite{ parascandolo2016recurrent,hayashi2017duration}.
Convolutional recurrent neural networks (CRNNs)~\cite{cakir2017convolutional, adavanne2017sound} have become a strong baseline for neural approaches.
More recently, self-attention-based models including Transformer~\cite{miyazaki2020weakly,kong2020sound,moritz2020all} and Conformer~\cite{miyazaki2020conformer} have shown significant improvement in AED performances.

These conventional methods focus mainly on \red{structures of the feature extractor} for modeling the characteristics of acoustic events and less on the part of the classifier design.
For the classifier design, almost all methods use multiple binary classifiers consisting of a linear layer and a sigmoid function.
This combination maps input audio to the activities of multiple events.
While this is a straightforward way to achieve multi-label classification, the assumption of conditional independence of each event activity is placed behind it.
This classifier performs estimation independently of other events, and there is no explicit interaction among the events.
However, sounds in the real world have dependencies on each other.
For example, `people speaking' is likely to occur with `people walking,' and `break squeaking' sounds should accompany the `car.'
Therefore, it is an inappropriate assumption that each event occurs independently.

Some AED methods focus on the classifier design and the co-occurrence of events.
Imoto {\it{et al}}.~\cite{imoto2019sound} modeled the co-occurrence relationship of events as a graph and used it as a constraint for training the model parameters.
They extracted co-occurrence relationships only as statistics in a specific length, such as within audio clips, and the constraint is used only during training.
Drossos {\it{et al}}.~\cite{drossos2019language} proposed an AED based on sequence-to-sequence modeling that takes into account the temporal context of acoustic events.
However, this method only models the input's temporal dependence, and the classifier itself is still the multiple binary classifiers.
Therefore, the assumption of conditional independence \red{still} remains in both cases.

In the machine learning field, the problem of label-dependencies regarding multi-label classification has been well studied~\cite{zhang2013review}.
For example, a classifier based on probabilistic chain rules~\cite{read2011classifier}, Bayesian approach to find label-dependencies has been proposed.
A neural network based method has also been proposed and shown its effectiveness in image classification~\cite{wang2017multi}, text classification~\cite{chang2020taming,gong2020hierarchical}, and the audio tagging task~\cite{cheng2010bayes}.
Also, in the speech field, 
multi-speaker speech separation, speech recognition, and speaker diarization with the chain rule~\cite{fujita2020neural,  shi2020sequence} have been proposed.
It demonstrates better performance than the conventional method, which deals with multiple speakers as conditionally.

This paper proposes acoustic event detection with classifier chains based on the probabilistic chain rule to handle the dependencies among events.
The proposed classifier chains consist of a gated recurrent unit and perform iterative \red{binary detection} of multiple events one by one.
In each iteration, one event's activity is estimated, and the estimated activity is used to condition the classifier chains for the next iteration.
The proposed method can handle the interdependence among events upon classification, \red{while} the conventional multiple binary classifiers has placed an assumption of conditional independence.
\begin{figure}
    \centering
    \includegraphics[width=0.8\linewidth]{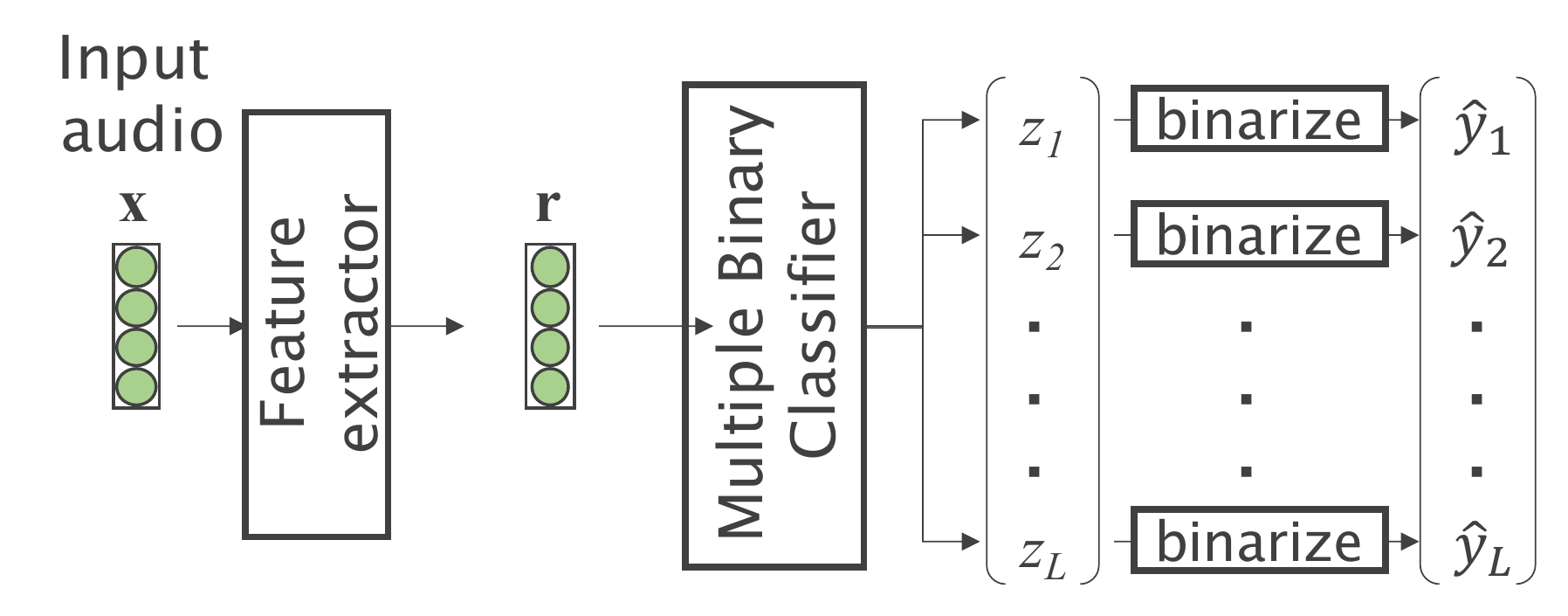} 
    \caption{General architecture of the conventional AED methods. The multiple binary classifiers estimate the activity for each event independently and there is no interaction among the events.}
    \label{fig:conv}
\end{figure}

\section{AED with classifier chains}
\subsection{Probabilistic formulation of AED and the conventional method}
Let us consider AED with $L$ event classes.
AED is a multi-label classification problem on a set of labels $\mathcal{L}$ that identifies what acoustic events are active in the input audio.
It can be formulated as a problem of estimating a subset of event labels $\hat{\mathcal{Y}} \subseteq \mathcal{L}$ from given input audio ${\bf{X}}\in\mathbb{R}^{T\times F}$:
\begin{align}
    \label{subset}
    \hat{\mathcal{Y}}= \underset{\mathcal{Y} \subseteq \mathcal{L}}{\text{argmax}} P( \mathcal{Y} \mid {\bf{X}}),
\end{align}
where ${\bf{X}} = \left[{\bf{x}}_1,...,{\bf{x}}_T\right]$ is a $F$ dimensional audio feature sequence of the input audio with $T$ temporal frames.
The label-subset $\mathcal{Y}$ is typically represented by an $L$-dimensional multi-hot vector ${\bf{y}}\in\{0,1\}^L$ that $\{0,1\}$ indicate activity \{inactive, active\} of each event class. 
Using this multi-hot activity representation, Eq. (1) can be rewritten as the following joint probability:
\begin{align}
    \label{subset}
    \hat{\bf{y}} &= \underset{{\bf{y}}\in\{0,1\}^L }{\text{argmax}} P( {\bf{y}} \mid {\bf{X}}) \\
                   &= \underset{y_1,...,y_L}{\text{argmax}} P( y_1,...,y_L \mid {\bf{X}}).
\end{align}
Figure~\ref{fig:conv} illustrates a general architecture of AED methods. 
The audio feature sequence ${\bf{X}}$ is transformed to a latent representation ${\bf{R}} = \left[{\bf{r}}_1,...,{\bf{r}}_T\right]$ by the feature extractor, where ${\bf{r}}\in\mathbb{R}^D$ is a $D$-dimensional latent representation.
The structure of the feature extractor is generally designed with neural networks, such as LSTMs~\cite{ parascandolo2016recurrent,hayashi2017duration}, CRNNs~\cite{cakir2017convolutional, adavanne2017sound}, and self-attention~\cite{miyazaki2020weakly,miyazaki2020conformer}.
The latent representation ${\bf{r}}_t$ at each temporal frame $t$ is fed into \red{any} classifier and obtain score vector ${\bf{z}}_t\in(0,1)^{L}$ whose elements represent the activity-score of the corresponding event class.
Then, each element of ${\bf{z}}_t$ is binarized with an appropriate threshold 
and resulting in an estimated activity of each event set $\hat{{\bf{y}}}_t$.

Almost all conventional methods employ multiple binary classifiers that consist of a linear layer and a sigmoid function.
Here, let ${\bf{W}}\in\mathbb{R}^{L\times{D}}$ and ${\bf{b}}\in{\mathbb{R}}^L$ denote the weight and bias parameters of the linear layer, respectively.
The multiple binary classifiers are written as follows:
\begin{align}
    \bf{z} = \sigma(\bf{Wr+b}), \\\label{cl}
    z_i= \sigma({\bf{w}}_i{\bf{r}}+b_i), 
\end{align}
where, ${\bf{w}}_i$ is $i$-th row vector in $\bf{W}$, $b_i$ is $i$-th elements of $\bf{b}$ and $\sigma(\cdot)$ represents the sigmoid function.
As can be seen from Eq.~(\ref{cl}), 
the estimation of the $i$-th class score $z_i$ depending only on the latent representation $\bf{r}$ and the $i$-th weight ${\bf{w}}_i$, and no information \red{on} the other event classes are involved.
Each event score and activity are estimated class-independently, depending only on the latent representation of the input audio.
In other words, there is the conditional independence assumption for each event in the conventional method.
\red{From the view point, } conventional methods approximate the joint probability of event classes in Eq.~(\ref{subset}) by the product of probabilities with class-independent:
\begin{align}
    \label{independent}
    P( {\bf{y}}\mid {\bf{X}}) &= P( y_1,...,y_L\mid {\bf{X}})\\
                            &=\Pi_{i=1}^{L} P( y_i \mid \hbox{\sout{$y_{i-1},...,y_1$}}, {\bf{X}}). 
\end{align}
\begin{figure}
    \centering
    \includegraphics[width=0.9\linewidth]{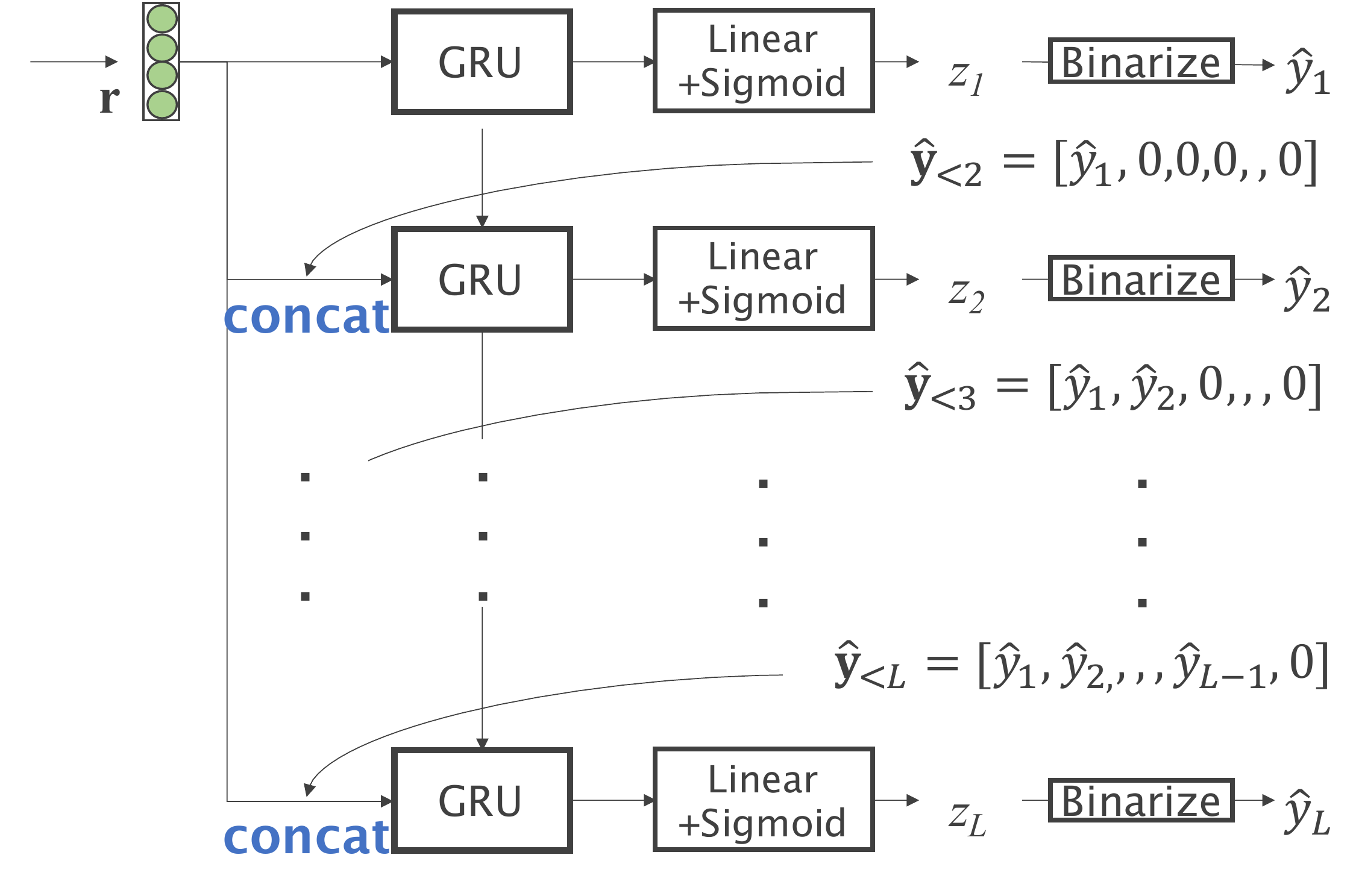} 
    \caption{The proposed classifier chains for AED. Each event activity is estimated iteratively and the estimation result in the previous iteration is used for conditioning next estimation.}
    \label{fig:chain}
\end{figure}

\subsection{Proposed AED with classifier chains}
The proposed method introduces classifier chains for AED, designed without conditional independence assumption based on the probabilistic chain rule.
The classifier of the proposed method iteratively estimates $y_i$ not only using the latent representation $\bf{r}$ but also conditioning by the estimated active events $\{\hat{y}_1,...,\hat{y}_{i-1}\}$ in the previous iterations.
Therefore, the classification of the proposed method is performed assuming the following joint probability:
\begin{align}
    \label{proposed}
    P( {\bf{y}} \mid {\bf{X}})  &=\Pi_{i=1}^{L} P( y_i \mid y_1,..., y_{i-1}, {\bf{X}}). 
\end{align}
For the multi-label classification, it is important to approximate the joint probability of multi-class in Eq. (3).
The conventional methods make the assumption of conditional independence on each class and approximate Eq. (3) with the product of the probability of each class in Eq. (7).
In contrast to the conventional method, the proposed classifier does not make the conditional independence assumption, and \red{is equivalent to} Eq. (3) based on the probabilistic chain rule as Eq. (8) 
so that the dependency among events can be modeled.
\red{Here, the order of the classes is an essential key in the chain rule. The order used in this paper and its impact on performance will be described in detail in the following sections.}

Figure~\ref{fig:chain} illustrates the proposed classifier chains.
The proposed method constructs classifier chains by iteratively estimating each event's activity using gated recurrent units (GRU).
First, the latent representation ${\bf{r}}$ is extracted from input audio ${\bf{x}}$ using a feature extractor.
For the feature extractor, any structure can be used.
In this paper, we employ the CRNN-based feature extractor, which is commonly used as a popular and strong baseline~\cite{cakir2017convolutional, adavanne2017sound,Turpault2020a}:
\begin{align}
    {\bf{R}} = \text{CRNN}^{(F\rightarrow D)}({\bf{X}}) \in \mathbb{R}^D.
\end{align}
The extracted latent representation ${\bf{R}} = \left[{\bf{r}}_1,...,{\bf{r}}_T\right]$ is fed into the classifier chain and event activities $\{\hat{y}_1,...,\hat{y}_L\}$ are estimated iteratively.
When estimating $\hat{y}_i$, the proposed method uses the latent representationn ${\bf{r}}$ of input audio and the estimated active event information $\hat{\bf{y}}_{<i} \in\{0,1\}^L$ in the previous iterations.
\red{The active event information $\hat{\bf{y}}_{<i}$ is represented as a multi-hot vector with elements which correspond to detected events are one, and the others are 0.}
Note that $\hat{\bf{y}}_{<i}$ has the fixed dimention $L$ for each iteration $i$ and elements with indices greater than $i$ are padded with zeros.
For example, suppose the third and fourth event classes have already been estimated as active in the previous iterations.
In that case, the third and fourth elements of $\hat{\bf{y}}_{<i}$ have 1, and others are 0, and it is used at the next iteration.
Using ${\bf{r}}$ and $\hat{\bf{y}}_{<i}$ , the estimation of $\hat{y}_i$ is as follows:
\begin{align}
    {\bf{r}}_i^{\prime} &= \text{Concatenate} \left({\bf{r}}, \hat{\bf{y}}_{<i}\right)  \in \mathbb{R}^{(D+L)}\\
    z_i &= \sigma\left(\text{Linear}^{(D+L \rightarrow 1)} ({\bf{r}}_i^{\prime})\right)\\
    \hat{y}_{i} &= \begin{cases}
    1 & z_{i}>\epsilon_i \\
    0 & \text{otherwise},
    \end{cases}
\end{align}
where $\epsilon_i$ is a threshold for binarizing $z_i$.
For the initial conditioning $\hat{\bf{y}}_{<0}$ is set to zero.

\subsection{The order of the classes for classifier chains}
The order of the classes is critical for classifier chains that iteratively identify each class.
In the chain rule-based speaker diarization~\cite{fujita2020neural}, 
the speaker's order for the chain rule is determined beforehand, verifying every speaker's permutation, and then the classifier chains are constructed.
However, the permutation of classes is a factorial of the number of classes.
Even if 10-class AED, as one of the typical AED task settings, there are 3,628,800 permutation patterns, and it is unrealistic to evaluate them all.
The proposed method, therefore, takes three approaches to set up the order of the classes.
\begin{description}
   \item[Higher/Lower F1 order]\mbox{}\\
   Orders based on event-wise f1-scores obtained preliminary experiment by the baseline method.
   In this paper, the order sets up two patterns: the order of the highest performance (Higher F1) and the lowest performance (Lower F1).
   \item[Higher/Lower Frequency order]\mbox{}\\
   Orders of the frequency of occurrence of each class in the training data set.
   \item[Random order]\mbox{}\\
   Orders determined by random permutations of the class index. In this paper, we investigated multiple random values.
\end{description}

\begin{table}[t]
\caption{Network architecture of the proposed method.}
\centering \vspace{-2mm}
\begin{tabular}{c} 
\toprule
{\bf Feature Extractor} \\  \toprule
Input: ($T$=512 frames, $F$=64 bins)\\ \midrule
2D CNN filter=(3, 3, 64) \\ 
Batch normalization, Pooling = (1, 4) \\  \midrule
2D CNN filter=(3, 3, 64) \\ 
Batch normalization, Pooling = (1, 4) \\  \midrule
2D CNN filter=(3, 3, 64) \\ 
Batch normalization, Pooling = (1, 4) \\  \midrule
Bi-GRU, Unit size= 62 \\ 
 \bottomrule
 \toprule
{\bf Classifier Chains} \\  \toprule
GRU, Unit size= 124 \\ 
Linear layer + Sigmoid \\ \midrule
Output shape = ($T$=512 frames, $L$= \# of classes$^*$) \\ 
\bottomrule
* 10 for URBAN-SED and 6 for TUT Sound Events 2017
\end{tabular}
\label{tab:net}
\end{table}
\section{Experiments}
\subsection{Datasets}
We conducted the experiments using two datasets; a synthetic dataset, URBAN-SED\cite{salamon2017scaper}, and a real-recording dataset, TUT Sound Events 2017~\cite{Mesaros2016_EUSIPCO}.
The URBAN-SED dataset is a synthetic dataset consisting of 10 acoustic event classes with 10,000 audio clips and annotations generated using the scaper library\cite{salamon2017scaper}.
The specification of each audio clip is a single channel, 44,100Hz, and 16-bit WAV format.
The files were split into the training set of 6,000, the validation set of 2,000, and the test set of 2,000.

TUT Sound Events 2017 dataset~\cite{Mesaros2016_EUSIPCO} is a real-recording dataset consisting of 6 event classes with manually annotated labels.
Each audio clip is recorded in a single acoustic scene, street, with a 44.1 kHz sampling rate and 24-bit resolution.

Each event in the URBAN-SED dataset is randomly synthesized to the background sound, and the frequency of occurrence and overlap of each event is artificially set.
On the other hand, TUT Sound Events 2017 dataset reflects the dependency among events in the real world.
Using these two different types of datasets, we can measure the effectiveness of the proposed chain-classifier in capturing event dependencies.

\subsection{Experimental conditions}
For the neural network input, every audio clip was transformed into a log-mel spectrogram of dimension $F=64$, a chunk of temporal frames $T=512$ with 20 ms window length and 10 ms hop length.
The network architecture of the proposed method is shown in Table~\ref{tab:net}.
2D CNN filter $= (A,B,C)$ denotes a 2D CNN layer with $(A, B)$ filter size and $C$ channels.
Pooling= $(a, b)$ denotes the max pooling operation on $a$ temporal frames and $b$ bins on the frequency axis.
The mini-batch size was 32, and the number of epochs was 100.
Adam~\cite{kingma2014adam} with learning rate 0.001 was used as the stochastic optimization method.
The activity-score threshold $\epsilon_i$ in Eq. (12) was optimized for the validation set.

The evaluation metric is frame-based and segment-based (1 sec.) macro f1-score
~\cite{Mesaros2016_MDPI}, which is the harmonic mean of recalls and precisions of frame/segment-wise classification results. 
The f1-scores were calculated for each event class, and the average of those f1-scores has been used.

For the comparison, we evaluated the CRNN with the multiple binary classifiers as ``baseline'', which is the strong and widely used baseline~\cite{cakir2017convolutional, adavanne2017sound,Turpault2020a}.
The difference with the proposed method was whether the classifier design is the classifier chains or the multiple binary classifier.
All other conditions were set up to be common.
In addition, for a more valid evaluation, we also evaluated a method with a CRNN-based feature extractor and a GRU classifier.
It is exactly the same architecture as the proposed method, only without the chain rule, i.e., without conditioning path in Figure~2.

As a validation of the proposed method, we trained the classifier chains using five class orders, as described in Section 2.3.
The five kinds of orders were, 
{\it{Higher/Lower F1 order}}, {\it{Higher/Lower Freq. order}} and {\it{Random order}}.

\begin{table}[t]
  \caption{AED performances of each classifier. All classifiers use CRNN for feature extractor.}
  \label{result_urban}
  \vspace{-2mm}
  \centering
\begin{tabular}{lcc}\hline
                           \multicolumn{3}{c}{URBAN-SED}                 \\ \hline
                          & Frame-based & Seg.-based \\\hline
Baseline                    & 0.612                & 0.625                  \\
+GRU (w/o Chain)           & 0.624                & 0.635                  \\
+\textbf{Chain (proposed)} & \textbf{0.631}       & \textbf{0.647}         \\ \hline
\end{tabular}

  \vspace{0mm} \vspace{0mm} \vspace{0mm}
  \centering
\begin{tabular}{lcc}\hline
                           \multicolumn{3}{c}{TUT Sound Event 2017}      \\ \hline
                          & Frame-based & Seg.-based \\ \hline
Baseline                    & 0.358                & 0.375                  \\
+GRU (w/o Chain)           & 0.375                & 0.395                  \\
+\textbf{Chain (proposed)} & \textbf{0.411}       & \textbf{0.426}         \\ \hline
\end{tabular}
\end{table}

\begin{table}[t]
  \caption{Effect of the chain-order on the frame-based F1-score. For both datasets, the higher the F1 order shows the best performance.
The averages for all orders are shown with their standard deviations.}
  \label{result_order}
  \vspace{-2mm}
  \centering
\begin{tabular}{lcc}
\hline
             & TUT Sound Events 2017         & URBAN-SED  \\ \hline\hline
Higher F1    & {\bf{0.411}}    & {\bf{0.631}}    \\
Lower F1     & 0.358    & 0.624    \\
Higher Freq. & 0.367      & 0.630    \\
Lower Freq.  & 0.368    & 0.630    \\
Random 1     & 0.364   & 0.627    \\
Random 2     & 0.382    & 0.624    \\
Random 3     & 0.355    & 0.628    \\
Random 4     & 0.380     & 0.629    \\
Random 5     & 0.385    & 0.624    \\ \hline\hline
Average      & 0.374$\pm{0.017}$ & 0.628$\pm{0.003}$ \\ \hline
\end{tabular}
\end{table}

\begin{table}[t]
\centering
\caption{Frame-based F1-scores for each event on the TUT Sound Events~2017 dataset. The event classes are listed in the Higher F1 order and the arrows indicate the order of the chain.}
\label{tab:my-table}
\vspace{-2mm}
\begin{tabular}{lcrlrl}
\hline
                 & \multirow{2}{*}{Baseline} & \multicolumn{2}{c}{Chain}                                         & \multicolumn{2}{c}{Chain}   \\
                 &                           & \multicolumn{2}{c}{Higher F1}                                         & \multicolumn{2}{c}{Lower F1}                                               \\ \hline
brakes squeaking              & \textbf{0.611}            & 0.573          & \multirow{6}{*}{\begin{tabular}[c]{@{}r@{}}$\shita$\\$\shita$\\$\shita$\\$\shita$\\$\shita$ \end{tabular}} & 0.532 & \multirow{6}{*}{\begin{tabular}[c]{@{}l@{}}$\ue$\\$\ue$\\$\ue$\\$\ue$\\$\ue$\end{tabular}} \\
car & 0.538                     & \textbf{0.583} &                                                      & 0.568 &                                                                    \\
people walking   & 0.487                     & \textbf{0.531} &                                                      & 0.527 &                                                                    \\
large vehicle    & \textbf{0.336}            & 0.324          &                                                      & 0.319 &                                                                    \\
people speaking  & 0.158                     & \textbf{0.380} &                                                      & 0.203 &                                                                    \\
children         & 0.017                     & \textbf{0.073} &                                                      & 0.000 &                                                                    \\ \hline \hline
\textbf{average} & 0.358                     & \textbf{0.411} & \textbf{}                                            & 0.358 &                                                                    \\ \hline
\end{tabular}
\end{table}

\subsection{Evaluation results}
Table~\ref{result_urban} shows experimental results of comparison among the baseline CRNN with the multiple binary classifiers (baseline), baseline with a GRU classifier (+GRU), and the proposed method using the classifier chains (+Chain). The chain order in these results is the Higher F1 order.
\red{In both datasets, the proposed method showed the highest performance. Comparing the GRU with the baseline, we can see the improvement due to the classifier architecture, and the proposed method further improves the performance.}
On TUT Sound Events 2017, the proposed method showed an improvement of 3.1\% on URBAN-SED and 14.8\% compared to the baseline.
The effect of the proposed method on URBAN-SED was limited because URBAN-SED is an artificially generated dataset with almost uniform dependency among events.
On the other hand, the impact of the proposed method on TUT Sound Events, which is real-recording data, was remarkable. 
This indicates that the proposed method can adequately capture the real-world dependencies among events and contributes to the classification.

\subsection{Impact of the chain order}
Table~\ref{result_order} shows f1-scores of each class-order for both dataset.
Among all class order settings, the Higher F1 order showed the best performance.
This is a convincing result, as the easiest class is classified first and then used to condition the subsequent classes.
\red{However, the performance was slightly worse than the baseline in some orders, such as Random 3 of TUT. Moreover, the average performance of all orders' results was almost equal to that of the GRU classifier. 
Therefore, the (inappropriate) choice of the order may harm the performance.
Particularly, for the real-recording dataset TUT Sound Events 2017, the effect of the order on performance was significant.
The performance with Random order 3 is 13.6\% lower than the best performance with the Higher F1 order.
For the synthetic dataset, URBAN-SED, there was little difference in performance by order, and the standard deviation of performance for each order was minimal compared to TUT Sound Events 2017.
This result also indicates that the effect of the proposed method is not significant for synthetic data sets where the dependency among classes is little or no class dependency.}

\subsection{Event-wise AED performance}
Table~\ref{tab:my-table} shows event-wise AED performance.
While some classes show degradation, the latter classes showed significant improvement.
For example, in `brakes squeaking,' the degradation was 6.3\%, and in `large vehicle,' the degradation was 3.5\%.
The class `brakes squeaking'  was the first class of the chain and classified without any conditioning, so its performance is considered to have worsened.
On the other hand, there were remarkable improvements in the class `people speaking' and class `children.'
This is due to the impact of conditioning became apparent as the chain progressed, leading to the remarkable improvement.
The Lower F1 order did not classify the `children' that were difficult to classify by the baseline.
Furthermore, these results also show that when a chain is started with a class that is difficult to classify, the overall performance shows degradation.
\red{
Since there are large errors in the estimated activity of the challenging class, the errors accumulate with each iteration leading to degradation of the performance in subsequent iterations
}.

\red{These results show that the proposed method is very powerful, but the performance is highly dependent on the choice of class order, which is consistent with the findings of speaker diarization~\cite{fujita2020neural,  shi2020sequence}.}

\section{Conclusion}\label{Sec:5}
This paper proposed acoustic event detection with the classifier chain.
The proposed classifier chains consist of GRU and iteratively detect multiple events.
\red{
In the experiments, the proposed method demonstrated its powerful performance comparing to the multiple binary classifiers.
The experimental results also showed the importance of the class order of the chain, which has a significant impact on the performance of the proposed method.
Future directions include an extension of the proposed method to weakly supervised training and investigating the effect of an order for datasets with a large number of classes, such as AudioSet~\cite{45857}.}

\ninept
\bibliographystyle{IEEEbib}
\bibliography{refs}

\end{document}